\documentclass[a4paper,12pt]{article}

\usepackage{ifpdf}

\newif\ifpdf
\ifx\pdfoutput\undefined
  \pdffalse
\else
  \pdfoutput=1
  \pdftrue
\fi

\RequirePackage{xspace} %
\RequirePackage{subfigure} %
\RequirePackage[centertags]{amsmath} %
\RequirePackage{amssymb}
\RequirePackage{wrapfig} %
\RequirePackage{calc} %
\RequirePackage{ifthen}
\RequirePackage{tabularx} %
\RequirePackage{flafter} %
\RequirePackage{fancyhdr} %

\ifpdf
  \RequirePackage[pdftex]{color}%
  \RequirePackage{colortbl}%
  \RequirePackage{array}%
  \RequirePackage[pdftex]{graphicx}

  \RequirePackage[ pdftex, plainpages = false, pdfpagelabels,
                 pdfpagelayout = useoutlines,
                 bookmarks,
                 breaklinks = true,
                 linktocpage,
                 pagebackref,                      
                 colorlinks = true,
                 linkcolor = blue,
                 urlcolor  = blue,
                 citecolor = blue,
                 anchorcolor = blue,
                 hyperindex = true,
                 hyperfigures
                 ]{hyperref}

\else
  \RequirePackage{color}
  \RequirePackage{colortbl}
   \RequirePackage{array}
  \RequirePackage[dvips]{graphicx}
  \RequirePackage{hyperref}
  \usepackage{rotating}
\fi

\usepackage{makeidx} 
\usepackage{setspace} 
\usepackage{rotating} 
\usepackage{ecltree}
\usepackage{epic}
\usepackage{supertabular}  
\usepackage{color}
\usepackage{exscale}
\usepackage{fontenc}
\usepackage{ifthen}
\usepackage{latexsym}
\usepackage{makeidx}
\usepackage{syntonly}
\usepackage{inputenc}
\usepackage{graphicx}
\usepackage{setspace}
\usepackage{caption2}
\usepackage[english]{babel}
\usepackage[square, comma,numbers,sort&compress]{natbib}
\usepackage{hypernat}
\usepackage{boxedminipage}
\usepackage{framed}
\usepackage{longtable}
\usepackage[all]{hypcap}    
\usepackage{algorithm2e}
\usepackage{algorithmic}
\usepackage{lscape}
\usepackage{pdflscape}

\newcommand{\note}[1]{\marginpar[left]{\singlespace \tiny #1}}

\newcommand{\HB}{Herschel-Bulkley}

\newcommand{\codi}    {converging-diverging} %
\renewcommand{\sectionmark}[1]%
      {\markright{\thesection\ #1}} 

\renewcommand{\note}[1]{}

\doublespace 

\title
{ %
\vspace*{3.0cm} \LARGE{\bf Methods for Calculating the Pressure Field in the Tube Flow} \vspace*{4.0cm} \\
}

\author{Taha Sochi\footnote{University College London, Department of Physics \& Astronomy, Gower Street, London, WC1E 6BT.
Email: t.sochi@ucl.ac.uk.} \vspace*{5.0cm}}

\begin{document}

\maketitle %
\pagenumbering{arabic}

\newpage
\phantomsection \addcontentsline{toc}{section}{Contents} %
\tableofcontents

\newpage
\phantomsection \addcontentsline{toc}{section}{Abstract} \noindent
{\noindent \LARGE \bf Abstract} \vspace{0.5cm}\\
\noindent
In this paper we outline methods for calculating the pressure field inside flow conduits in the
one-dimensional flow models where the pressure is dependent on the axial coordinate only. The
investigation is general with regard to the tube mechanical properties (rigid or distensible), and
with regard to the cross sectional variation along the tube length (constant or variable). The
investigation is also general with respect to the fluid rheology as being Newtonian or
non-Newtonian.

\vspace{0.5cm}


\pagestyle{headings} %
\addtolength{\headheight}{+1.6pt}
\lhead[{Chapter \thechapter \thepage}]%
      {{\bfseries\rightmark}}
\rhead[{\bfseries\leftmark}]%
     {{\bfseries\thepage}} 
\headsep = 1.0cm               

\newpage
\section{Introduction}\label{Introduction}

In the flow of Newtonian and non-Newtonian fluids through conduits of different geometries and
various wall mechanical properties the relation that is usually in demand is the volumetric flow
rate as a function of the pressure at the two boundaries with no need for a detailed knowledge
about the pressure field inside the conduit, because in the case of rigid tubes the flow rate
generally depends on the difference between the pressure at the inlet and outlet boundaries, while
for the distensible tubes the flow rate is dependent on the actual values of the pressure at the
two boundaries \cite{SochiElastic2013}. Moreover, in the common cases of purely-viscous laminar
axi-symmetric flow in rigid conduits with a constant radius along their length the pressure field
can be obtained by simple linear interpolation and symmetric arguments with no need for elaborate
calculations. Therefore, in most cases the pressure field inside the conduit is either not of
primary interest to the investigator or it can be obtained with a minimal effort.

However, in some other circumstances the pressure as a function of the luminal space coordinates
may be needed and it cannot be obtained from simple arguments due to the complexity of flow
situation arising for instance from a history dependent non-Newtonian fluid or tube wall
distensibility or complicated tube geometry. For example, in assessing the durability and
resistance of a pipeline in the risk assessment analysis or in the case of predicting the yield
point of a yield-stress fluid in a distensible tube \cite{SochiElasticYield2013} detailed knowledge
of more complex pressure fields is required. In such circumstances, elaborate analytical and
numerical techniques are needed to obtain the pressure field.

Various methods have been used in the past to calculate the pressure field in the flow conduits.
Many of these methods are based on the use of analytical techniques and apply to special cases of
flow, fluid and mechanical and geometrical characteristics of conduit. Numerical discretization
techniques such as finite element and finite difference methods have also been used for finding the
pressure fields although in most circumstances the pressure field may be regarded as a byproduct of
the primary computational objectives.

The literature of fluid dynamics contains numerous studies dedicated to the investigation and
analysis of pressure fields inside flow conduits and storage vessels. For example, Miekisz
\cite{Miekisz1963} investigated the temporal and spatial dependency of the pressure inside an
elastic tube in the context of hemodynamic viscous flow using a linear telegraph equation with the
employment of analytical mathematical techniques. Oka \cite{Oka1973} studied the pressure field in
a tapered tube and developed a formula for calculating the pressure gradient as a function of the
axial coordinate for the viscous flow of non-Newtonian fluids using power law, Bingham and Casson
models. In a series of studies, Degasperi and coworkers \cite{DegasperiVMT2004, DegasperiMTV2004,
DegasperiB2004, DegasperiMTV2005} investigated the pressure field distribution in tubes with
different geometries under transient and steady state conditions in the context of injection and
outgassing of vacuum devices using analytical and numerical techniques. Naili and Thiriet
\cite{NailiT2005} studied the pressure field in straight rigid tubes with a uniform cross section
in the axial direction under Newtonian incompressible flow conditions using five different cross
sectional shapes. There are many other studies (e.g. \cite{BarnardHTV1966, LahbabiC1986,
LetelierSC2002, MasulehP2004, Tripathi2011, DescovichPMSW2013, TripathiB2013}) which are not
dedicated to computing the pressure although the methods and results can be used to obtain the
pressure field. There are also numerous studies related to the pressure wave propagation in
distensible tubes and blood vessels (e.g. \cite{Paidoussis1965, MisraS1989, WangCCHW1997,
BerkoukCL2003, CarpenterBL2003, Kizilova2006, KitawakiS2009, SaitoIMWAe2011b}); a subject which is
not included in the scope of the present paper.

In the one-dimensional flow models, where the flow is assumed to have axial dependency only in the
$z$ direction of the cylindrical coordinate system, the pressure is well defined at each axial
point with no dependency on the azimuthal angle or the radial distance. For rigid tubes with
constant radius along their axial length the pressure is generally a linear function of the axial
coordinate, i.e. constant pressure gradient, and hence the axial pressure can be trivially
evaluated from the given boundary conditions through linear interpolation. However, for rigid tubes
with varying cross section along their axial direction and for distensible tubes in general the
pressure is not a linear function of the axial coordinate and hence the axial pressure cannot be
computed from the given boundary conditions using a simple linear interpolation formula. A similar
situation may also arise in the flow of history dependent (i.e. viscoelastic and
thixotropic/rheopectic) non-Newtonian fluids in rigid tubes with a constant radius due to the
possible change of rheology and flow profile along the tube length and hence the axial pressure
field may not be a linear function of the axial coordinate.

In the present investigation we outline some methods that have been used or can be used in the
calculation of the axial pressure field in the one-dimensional flow models. Although, the
investigation is general with regard to the type of fluid as Newtonian or non-Newtonian, and the
type of conduit i.e. with constant or variable cross section and rigid or distensible mechanical
wall characteristics, it does not extend to the case of history dependent fluids due to the
complexity of these problems and the lack of viable and sufficiently simple flow models that can be
used to analyze such flows. For the generality of discussion and results, we assume throughout this
investigation a laminar, steady state, incompressible, slip-free, one-dimensional, fully developed
flow with negligible body and inertial forces and minor entrance and exit effects although some of
these assumptions and conditions can be relaxed in some circumstances.

\section{Methods}

In this section we outline some of the methods that can be used for finding the axial pressure
field inside a tube containing a fluid and subject to a pressure gradient. As indicated already,
the tube is more general than being circular cylindrical with rigid walls and constant cross
section along its length under Newtonian flow conditions. However, there is no single method that
is sufficiently general to apply to all these variations of flow, fluid and conduit in all
circumstances as will be clarified in the forthcoming subsections.

\subsection{Solving Navier-Stokes Equations}

Using Navier-Stokes flow equations, which are based on the mass and linear momentum conservation
principles, is the most common way for describing the flow of fluids and obtaining its parameters
including the pressure field. One example of using the Navier-Stokes equations to obtain the axial
pressure is the case of the one-dimensional form of the Navier-Stokes flow in distensible tubes
where an implicitly defined axial pressure field can be obtained analytically (see
\cite{SochiElasticYield2013} and Equations \ref{NSAnaSolEq}). However there are many limitations on
the validity and direct use of the Navier-Stokes equations. One reason is the limiting assumptions
on which the Navier-Stokes flow is based; for example the fluid is assumed Newtonian and hence any
extension beyond this requires an extension to the Navier-Stokes model.

There are also many practical difficulties in obtaining analytical solutions from the Navier-Stokes
equations due to their non-linear nature associated potentially with other mathematical
difficulties originating for example from a complex flow geometry. So even if the Navier-Stokes
flow is valid in principle for describing the problem the subsequent mathematical difficulties may
prevent obtaining a solution. However, numerical methods can in most cases bridge such gaps and
hence numerical solutions can be obtained. Another difficulty is that even if an analytical
solution can be obtained it may not be possible to obtain an explicit functional form of the axial
pressure in terms of the axial coordinate and hence a numerical method is needed to obtain the
pressure field numerically from its implicit definition. However, obtaining a solution numerically
from an implicit form of an analytical expression is easy through the use of a simple numerical
solver based for instance on a bisection method.

\subsection{Using Flow Relation}

This method is based on the exploitation of an available relationship correlating the volumetric
flow rate to the two pressure boundary conditions at the inlet and outlet. An equation can then be
formed by using the boundary conditions and an arbitrary axial point with the elimination of the
flow rate by equating the flow rate in one part of the tube to the flow rate in the other part. For
example if we have the following general relation

\begin{equation}
Q=f(p_i,p_o,L)
\end{equation}
where $Q$ is the volumetric flow rate, $f$ is a function with a known form, $p_i$ and $p_o$ are the
inlet and outlet pressure respectively, and $L$ is the tube length, then we can form two equations
each one with one boundary condition and an arbitrary axial point, i.e.

\begin{equation}
Q=f(p_i,p_x,x) \hspace{1cm} \textrm{and} \hspace{1cm} Q=f(p_x,p_o,L-x)
\end{equation}
where $p_x$ is the axial pressure at an arbitrary axial coordinate $0<x<L$. Because the flow rate
is constant, due to the incompressible steady state assumptions, then by eliminating $Q$ between
the above two equations a single equation can be formed which links explicitly or implicitly $p_x$
to $p_i$ and $p_o$ and hence $p_x$ can be obtained either directly, if it is explicitly defined, or
by the use of a simple numerical solver based for instance on the bisection method if it is an
implicit function.

Alternatively, the value of the volumetric flow rate $Q$ is computed from the two given boundary
conditions and then substituted in the $p$-$Q$ relation with one of the given boundary conditions.
The second boundary condition is then replaced with the condition at an arbitrary axial point
$0<x<L$ and the resulting equation is solved, usually numerically, to obtain the unknown ``boundary
condition'' which is the axial pressure at the internal point with $x$ coordinate.

We clarify this with an example from the one-dimensional Navier-Stokes flow in elastic tubes where
a linear elastic relation between the axial pressure and the corresponding cross sectional area of
the following form is used to describe the mechanical response

\begin{equation}
p=\gamma\left(A-A_{o}\right)\label{pAEq1}
\end{equation}
where $p$ is the axial pressure, $\gamma$ is a stiffness coefficient, $A$ is the tube cross
sectional area at the axial pressure $p$, and $A_{o}$ is the reference area corresponding to the
reference pressure which, in this equation, is set to zero for convenience without affecting the
generality of the results.

For the distensibility mechanical relation of Equation \ref{pAEq1}, the following formula that
links the volumetric flow rate to the inlet and outlet boundary conditions is derived and validated
previously \cite{SochiElastic2013}

\begin{equation}\label{QElastic1}
Q=\frac{L-\sqrt{L^{2}-4\frac{\alpha}{\kappa}\ln\left(A_{ou}/A_{in}\right)\frac{\gamma}{3\kappa\rho}\left(A_{in}^{3}-A_{ou}^{3}\right)}}{2\frac{\alpha}{\kappa}\ln\left(A_{ou}/A_{in}\right)}
\end{equation}
where $\alpha$ is a correction factor for the axial momentum flux, $\kappa$ is a viscosity friction
coefficient, $\rho$ is the fluid mass density, and $A_{in}$ and $A_{ou}$ are the tube cross
sectional area at the inlet and outlet respectively. If this equation is solved for two given
boundary conditions to obtain a value for the volumetric flow rate $Q_v$, then the following
equation can be formed using the inlet boundary condition

\begin{equation}\label{QElasticin}
\frac{x-\sqrt{x^{2}-4\frac{\alpha}{\kappa}\ln\left(A_{x}/A_{in}\right)\frac{\gamma}{3\kappa\rho}\left(A_{in}^{3}-A_{x}^{3}\right)}}{2\frac{\alpha}{\kappa}\ln\left(A_{x}/A_{in}\right)}=Q_v
\end{equation}
where $A_{x}$ is the tube cross sectional area at an arbitrary axial point $0<x<L$. The latter
equation can then be solved numerically for $A_{x}$ and the pressure at $x$ is then obtained from
$A_{x}$ by using Equation \ref{pAEq1}. Alternatively, the outlet boundary condition can be used
with an arbitrary axial point representing an inlet boundary, that is

\begin{equation}\label{QElasticou}
\frac{(L-x)-\sqrt{(L-x)^{2}-4\frac{\alpha}{\kappa}\ln\left(A_{ou}/A_{x}\right)\frac{\gamma}{3\kappa\rho}\left(A_{x}^{3}-A_{ou}^{3}\right)}}{2\frac{\alpha}{\kappa}\ln\left(A_{ou}/A_{x}\right)}=Q_v
\end{equation}

The main requirement for the applicability of this method is that the characteristic flow equation
is indifferent to the axial coordinate due to the fact that the tube has a uniform reference cross
section, i.e. either rigid with a constant cross section or distensible with a constant cross
section under reference pressure.

One limitation of this method is the need for a pressure-flow rate relationship which usually
should be obtained analytically. Another limitation is the aforementioned requirement which
restricts the method to certain geometries as required by the $p$-$Q$ expression, and hence it
cannot be used for instance to find the pressure field in \codi\ geometries that have analytical
$p$-$Q$ relations \cite{SochiPower2011, SochiNavier2013}. Also, the method normally requires the
employment of a numerical solver because the resulted expression is usually implicit in its
definition of the axial pressure as a function of the given conditions; although only a simple
numerical solver, such as a bisection solver, is needed.

\subsection{Traditional Meshing Techniques}

Another method is the use of the traditional meshing techniques such as finite element and finite
difference \cite{SmithPH2002, FormaggiaLQ2003, SherwinFPP2003, ReymondMPRS2009,
SochiTechnical1D2013}. These methods are widely used for solving flow problems although they may
not be the best available option for finding the pressure field due partly to the approximations
and possible errors and bugs; moreover most of these methods are not easy to implement.

\subsection{Residual-Based Lubrication}

For any history independent characteristic flow, such as the flow of Newtonian fluids in elastic
tubes or the flow of non-Newtonian fluids of a \HB\ type into rigid tapered tubes, there is a
characteristic flow relation that links the volumetric flow rate through the tube to the two
pressure boundary conditions at the inlet and outlet. The residual-based lubrication method
exploits such a characteristic flow relation by discretizing the flow conduit into axially-divided
thin slices where the flow relation is assumed to equally apply to each one of these slices. These
slices either have the same width or have different widths but where the largest width is below a
predefined maximum limit.

For a conduit discretized into $(N-1)$ slices there are $N$ axial nodal points: two boundaries and
$(N-2)$ interiors. A system of $N$ simultaneous equations based on the two boundary conditions for
the boundary nodes and the conservation of flow rate, which is equivalent to the conservation of
mass for an incompressible flow, for the interior nodes is then formed and solved by a non-linear
solution scheme such as Newton-Raphson method subject to a predefined error tolerance. The result
of this solution is the axial pressure at the interior nodes, which defines the pressure field, and
the volumetric flow rate. More detailed description of this method can be obtained from
\cite{SochiPoreScaleElastic2013, SochiConDivElastic2013, SochiCoDiNonNewt2013}.

The residual-based lubrication method is the most general method for obtaining the pressure field
as compared to the other methods in terms of the fluid, i.e. Newtonian or non-Newtonian, and the
flow conduit, i.e. rigid or distensible, with a constant or variable cross section along its axial
direction. In fact the only requirement for the applicability of the residual-based lubrication
method is the availability of a characteristic flow relation that describes the flow on each slice,
where this relation can be analytical, empirical or even numerical \cite{SochiVariational2013}.
Therefore, the flow conduit is not required to have a constant shape or size in the axial
direction. In fact the residual-based method can equally apply to conduits with non-circular cross
sections. Moreover, it can apply in principle to conduits with more than one characteristic flow
relation where these different relations apply to different parts of the conduit, e.g. one apply to
a first part with circularly-shaped cross sections and another apply to a second part with
squarely-shaped cross sections. However, the residual-based lubrication method can not be applied
in some circumstances due to the lack of a characteristic flow relation; e.g. in most cases of flow
of non-Newtonian fluids through distensible conduits.

\begin{figure}[!h]
\centering{}
\includegraphics
[scale=0.75] {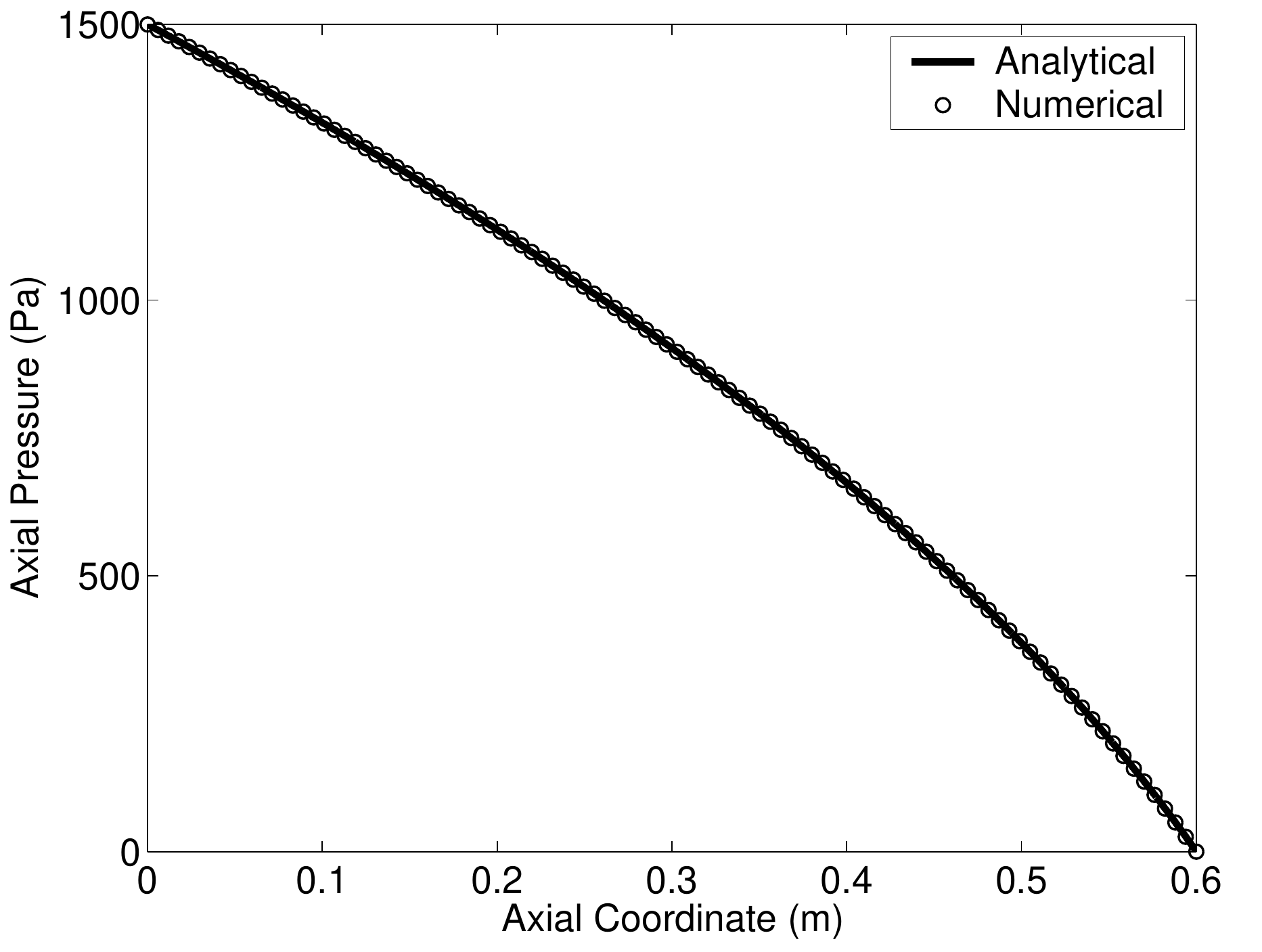} %
\caption{Comparison between the analytical solution given by Equation \ref{NSAnaSolEq} and the
numerical solution obtained from the residual-based lubrication method for the one-dimensional
Navier-Stokes flow in an elastic tube. The parameters for this example are: $A_o=0.0201062$~m$^2$,
$L=0.6$~m, $\alpha=1.25$, $\beta=10635$~Pa.m, $\rho=1100$~kg.m$^{-3}$, $\mu=0.035$~Pa.s,
$p_i=1000$~Pa, $p_o=0$~Pa, and $Q=0.101768$~m$^3$.s$^{-1}$.} \label{NSCompare}
\end{figure}

The residual-based lubrication method can provide very accurate solutions if a proper
discretization with a low error tolerance were employed. To demonstrate this fact, in Figure
\ref{NSCompare} we compare the analytical solution for the pressure field in a distensible tube
with a Navier-Stokes one-dimensional flow to the numerical solution as obtained from the
residual-based method. The analytical solution, which is derived in \cite{SochiElasticYield2013},
is given by

\begin{equation}\label{NSAnaSolEq}
x=\frac{2\alpha
Q\ln\left(\frac{\frac{A_{o}}{\beta}p+\sqrt{A_{o}}}{\frac{A_{o}}{\beta}p_{i}+\sqrt{A_{o}}}\right)}{\kappa}+\frac{\beta\left[\left(\frac{A_{o}}{\beta}p_{i}+\sqrt{A_{o}}\right)^{5}-\left(\frac{A_{o}}{\beta}p+\sqrt{A_{o}}\right)^{5}\right]}{5\rho\kappa
QA_{o}}
\end{equation}
where $\beta$ is a stiffness coefficient in the presumed non-linear pressure-area elastic relation,
while the other symbols are as defined previously. The analytical relation of Equation
\ref{NSAnaSolEq} is based on the following distensibility elastic relation, whose essence is that
the axial pressure is proportional to the corresponding radius change with a proportionality
stiffness factor scaled by the reference area, that is

\begin{equation}\label{pAEq2}
p=\frac{\beta}{A_{o}}\left(\sqrt{A}-\sqrt{A_{o}}\right)
\end{equation}

Other advantages of the residual-based lubrication method include relative ease of implementation,
reliable and fast convergence, and low computational costs in terms of memory and CPU time as
compared to equivalent methods like finite element.

\subsection{Special Methods}

As indicated early, many of the previous attempts to find the pressure field are based on special
mathematical and computational methods which are tailored for the given problem with the presumed
geometry and flow characteristics. A sample of these attempts and the employed methods have been
highlighted in the Introduction. It is out of the scope of the present paper to go through these
special methods.

\section{Conclusions}

In this study we outlined general methods that can be used to obtain the axial pressure as a
function of the axial coordinate for the Newtonian and non-Newtonian one-dimensional flow in the
rigid and distensible tubes with constant and variable cross sections along their length. As seen,
there is no single method that can be used in all cases although the residual-based lubrication
method is the most general one with respect to the type of fluid  (Newtonian or non-Newtonian), and
the type of tube (rigid or distensible, having a uniform cross section or axially varying cross
section). The residual-based method has also other advantages such as comparative ease of
implementation, low computational cost in terms of CPU time and memory, fast and reliable
convergence and highly-accurate solutions which compare in their accuracy to any existing or
potential analytical solutions if sufficiently fine discretization and low error tolerance are
used.

\clearpage
\phantomsection \addcontentsline{toc}{section}{Nomenclature} %
{\noindent \LARGE \bf Nomenclature} \vspace{0.5cm}

\begin{supertabular}{ll}
$\alpha$                &   correction factor for axial momentum flux \\
$\beta$                 &   stiffness coefficient in the non-linear pressure-area elastic relation \\
$\gamma$                &   stiffness coefficient in the linear pressure-area elastic relation \\
$\kappa$                &   viscosity friction coefficient \\
$\mu$                   &   fluid dynamic viscosity \\
$\rho$                  &   fluid mass density \\
\\
$A$                     &   cross sectional area of elastic tube at given axial pressure \\
$A_{in}$                &   cross sectional area of elastic tube at inlet \\
$A_o$                   &   cross sectional area of elastic tube at reference pressure \\
$A_{ou}$                &   cross sectional area of elastic tube at outlet \\
$A_x$                   &   cross sectional area of elastic tube at $x$ coordinate \\
$L$                     &   tube length \\
$p$                     &   pressure \\
$p_i$                &  inlet boundary pressure \\
$p_o$                &  outlet boundary pressure \\
$Q$                     &   volumetric flow rate \\
$Q_v$                   &   numeric value of volumetric flow rate \\
$x$                     &   tube axial coordinate \\

\end{supertabular}

\clearpage
\phantomsection \addcontentsline{toc}{section}{References} %
\bibliographystyle{unsrt}

\end{document}